\documentclass[aps,prd,preprint,amsmath,amssymb,longbibliography,nofootinbib,eqsecnum]{revtex4-2}

\usepackage{bm}
\usepackage{braket}
\usepackage{tensor}
\usepackage{mathtools}
\usepackage{amsmath}

\usepackage{graphicx}
\usepackage{subcaption}
\usepackage{multirow}
\usepackage{dcolumn}
\newcolumntype{d}[1]{D{.}{.}{#1}}

\usepackage{xcolor}
\definecolor{DarkBlue}{rgb}{0,0,0.4}
\definecolor{DarkGreen}{RGB}{0,100,0}

\usepackage{hyperref}
\colorlet{LinkColor}{DarkBlue}
\colorlet{CiteColor}{DarkBlue}
\colorlet{URLColor}{DarkBlue}
\hypersetup{%
    citecolor=CiteColor,
    colorlinks=true,
    linkcolor=LinkColor,
    urlcolor=URLColor,
}

\newcommand{\orcid}[1]{%
\begingroup
  \hypersetup{hidelinks}\href{https://orcid.org/#1}{\includegraphics[width=10pt]{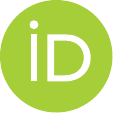}}
\endgroup}

\usepackage{comment}

\usepackage{soul} % Command: \st{}
\colorlet{CorrectionColor}{red}
\setstcolor{CorrectionColor}
\newcommand{\mathst}[1]%
{\bgroup\mathchoice
  {\sbox0{$\displaystyle{#1}$}%
    \usebox0\hspace{-\wd0}%
    \textcolor{red}{\rule[0.5\ht0-0.5\dp0-.5pt]{\wd0}{1pt}}}%
  {\sbox0{$\textstyle{#1}$}%
    \usebox0\hspace{-\wd0}%
    \textcolor{red}{\rule[0.5\ht0-0.5\dp0-.5pt]{\wd0}{1pt}}}%
  {\sbox0{$\scriptstyle{#1}$}%
    \usebox0\hspace{-\wd0}%
    \textcolor{red}{\rule[0.5\ht0-0.5\dp0-.5pt]{\wd0}{1pt}}}%
  {\sbox0{$\scriptscriptstyle{#1}$}%
    \usebox0\hspace{-\wd0}%
    \textcolor{red}{\rule[0.5\ht0-0.5\dp0-.5pt]{\wd0}{1pt}}}%
\egroup}

\begin{document}

\title{%
    Axion condensates in neutron stars and radial oscillation modes
}

\author{%
    Antonio Gómez-Bañón 
    \orcid{0009-0007-2546-5372}
}
\affiliation{Departament de Física, Universitat d’Alacant, 03690 Alicante, Spain}

\author{%
    Pantelis Pnigouras
    \orcid{0000-0003-1895-9431}
}
\affiliation{Departament de Física, Universitat d’Alacant, 03690 Alicante, Spain}

\author{%
    Jos\'e A. Pons 
    \orcid{0000-0003-1018-8126}
}
\affiliation{Departament de Física, Universitat d’Alacant, 03690 Alicante, Spain}

\date{\today}

\begin{abstract}
Light QCD axions, introduced to solve the strong \textit{CP} problem, may form condensates inside neutron stars, giving rise to a novel ground state of dense matter. We investigate how such axion condensates modify the equilibrium structure and radial oscillation spectrum of neutron stars. Using a realistic neutron star model with the BSk26 equation of state, and solving the coupled Tolman–Oppenheimer–Volkoff and Klein–Gordon equations together with a linear perturbation analysis, we find two distinct families of quasinormal modes: weakly damped fluid-dominated oscillations and highly damped axion modes. The coupling between the fluid and the axion field introduces axion-induced damping of radial oscillations, with decay timescales of order seconds for kHz axion masses. Modes with frequencies above the axion mass are strongly damped, while those below remain unaffected. Although neutron star radial oscillations are difficult to observe, our results suggest that extensions of this work can turn neutron star seismology into a novel probe of the axion properties.
\end{abstract}

\maketitle

\section{Introduction} \label{sec:Introduction}

Axions were originally introduced as a solution to the strong \textit{CP} problem in QCD \cite{Peccei_1977a,Peccei_1977b,Wilczek_1978,Weinberg_1978}. These hypothetical particles have since become increasingly popular in proposed extensions of the standard model. Due to their extremely small mass and weak interactions with ordinary matter, axions are expected to play a major role in the dark matter content of the Universe \cite{Abbott_1983,Dine_1983,Preskill_1983}. Current cosmological surveys provide some of the most stringent bounds on the axion properties \cite{Depta_2020,Langhoff_2022,Notari_2023,DiValentino_2023,Bianchini_2024}, but the potentially ubiquitous presence of axions in different scenarios makes them potentially relevant across a wide range of astrophysical contexts.

A variety of astrophysical systems display distinctive signatures under the assumption of the axion existence, allowing constraints to be placed on its mass and couplings to radiation and matter. Examples include axion-induced emission from main-sequence stars such as the Sun \cite{Raffelt_1986,Lazarus_1992,Moriyama_1998,Arik_2009}, supernova cooling \cite{Raffelt_1988, Turner_1988}, neutron star (NS) cooling \cite{Buschmann_2022,Gomez_2024}, axion-to-photon conversions in NS magnetic fields \cite{Sikivie_1983,Raffelt_1988b,Pshirkov_2009,Hook_2018b}, and black hole superradiant instabilities \cite{Arvanitaki_2015,Cardoso_2018,Davoudiasl_2019,Abbott_2022}.

In this work, we consider a particularly interesting scenario in which the axion field can condense inside NSs, giving rise to a new ground state (NGS) of nuclear matter \cite{Balkin_2024}. This becomes possible via the inclusion of an additional degree of freedom in the effective potential of the standard QCD axion, in the form of a small parameter $\epsilon$, effectively lowering the axion mass and allowing the interaction with baryonic matter to source an axion condensate inside the star \cite{Hook_2018}. This results in observable consequences that can place constraints on the axion model \cite{Gomez_2024, Balkin_2024}. For instance, in \cite{Gomez_2024}, it was shown that axion condensation modifies the exterior layers of NSs, making their envelopes thinner. This leads to anomalously fast cooling, which rules out a large portion of the axion parameter space (see Fig.~5 in \cite{Gomez_2024}). These findings were confirmed independently by \cite{Kumamoto_2025}, making the bounds even more restrictive.

As part of the ongoing efforts to constrain the axion hypothesis, in this paper we explore the effects of the axion condensate on the oscillation modes of NSs. Normal modes are an essential tool for probing NS interiors and can potentially offer abundant information about the largely unknown dense nuclear matter equation of state (EOS; see, e.g., \cite{McDermott_1988,Andersson_1998,Kokkotas_1999,Ferrari_2003,Benhar_2004,Doneva_2013,Suvorov_2018,Krueger_2020,Gittins_2025} for a selection from the extensive literature). Each piece of physics of the NS is associated with a family of modes (e.g., composition or entropy gradients \cite{McDermott_1983,Reisenegger_1992,Gaertig_2009,Counsell_2025}, superfluidity \cite{Passamonti_2012,Gualtieri_2014}, the presence of an elastic crust \cite{Vavoulidis_2007,Krueger_2015}, and magnetic fields \cite{Passamonti_2014,Gabler_2016}), leading to a very rich phenomenology across both the electromagnetic and the gravitational wave (GW) spectrum. Unsurprisingly, the presence of the axion field introduces a new family of modes and also influences the regular fluid modes.

As a first step, we investigate radial oscillation modes. Although usually considered uninformative in NSs, as they do not emit GWs, radial modes have been thoroughly studied in the context of relativistic stars for a wide range of EOSs \cite{Meltzer_1966,Chanmugam_1977,Vaeth_1992,Kokkotas_2001}. In the presence of the axion field, which couples to baryonic matter, an additional branch of radial perturbations emerges, damped due to axion emission. The behavior is akin to that of the spacetime $w$-modes \cite{Kokkotas_1986,Kokkotas_1992}, with the axion condensate in this context being analogous to the spacetime.

Even though NS radial modes remain difficult to observe, the essential physical mechanisms are expected to carry over to the nonradial mode case. This work provides both a foundation and a motivation on which future extensions can be based, suggesting that NS oscillation modes could potentially be used to probe axion properties.

We organize this paper as follows. First, in Sec.~\ref{sec:back}, we review the equilibrium structure of NSs with an axion condensate. In Sec.~\ref{sec:radosc}, we introduce the set of equations and boundary conditions describing the radial perturbations of the NS-axion system. In Sec. \ref{sec:simple}, we present a simplified model that captures the most relevant physics of the system. In Sec.~\ref{sec:res}, we present our results for the oscillation frequencies in a realistic NS, including their division into two families based on their damping timescales, along with a detailed discussion. Finally, we conclude with a brief summary in Sec.~\ref{sec:summary}. Unless otherwise indicated, throughout the paper we use natural units, in which $\hbar=c=1$.

\section{The background: a neutron star with an axion soul}\label{sec:back}

\subsection{Light QCD axions at finite density}

Light QCD axions are distinguished by a suppressed vacuum mass, commonly parametrized by a small dimensionless quantity $\epsilon$ that quantifies the degree of suppression. The corresponding vacuum potential takes the form \cite{DiVecchia_1980, di_Cortona_2016}
\begin{subequations}
\begin{align}
    V(a)&=\epsilon m_\pi^2 f_\pi^2\left[1-g(a)\right],
    \label{eq:axion potential}\\
    \shortintertext{with}
    g(a) &\equiv\sqrt{1-\beta\sin^2\left(\frac{a}{2 f_a}\right)}\label{eq:gfunc},
\end{align}    
\end{subequations}
where $a$ denotes the axion field, $f_a$ is the axion decay constant, $m_\pi$ and $f_\pi$ are the pion mass and decay constant, respectively, and $\beta\equiv4z/(1+z)^2\approx0.88$ (with $z\approx0.48$ denoting the up-to-down-quark mass ratio). Expanding Eq.~\eqref{eq:axion potential}, the axion mass may be identified as $m_a=\sqrt{\epsilon \beta} f_\pi m_\pi/2 f_a$. The significance of the parameter $\epsilon$ now becomes apparent--for a fixed $f_a$, it modifies the value of the axion mass, making it lighter than that of the standard QCD axion (which is recovered for $\epsilon=1$ \cite{di_Cortona_2016}) by a factor of $\sqrt{\epsilon}$.

At energies below the QCD scale, the Lagrangian acquires an interaction term of the form \cite{Ubaldi_2010, Hook_2018, Balkin_2020}
\begin{equation}
\label{eq:axion_nucleon_int}
\mathcal{L}_{\mathrm{int}}(a, n_s)=\sigma_N \Bar{N}N\left[1-g(a)\right],
\end{equation}
where $N=\left(p, n\right)^T$ is the nucleon field and $\sigma_N\approx50~\mathrm{MeV}$ is the nucleon sigma term \cite{Alarcon_2021}. In the presence of a nonvanishing baryon scalar density $n_s\equiv\langle\Bar{N}{N}\rangle$, the interaction term of the Lagrangian results in a modified axion potential $U$, given by
\begin{equation}
    U(a, n_s)\equiv V(a)-\mathcal{L}_{\mathrm{int}}(a, n_s)=\left(\epsilon m_\pi^2 f_\pi^2-\sigma_Nn_s\right) \left[1-g(a)\right].
    \label{eq:modified axion potential}
\end{equation}
Hence, for $n_s>n_c\equiv\epsilon m_\pi^2 f_\pi^2/\sigma_N$ the effective potential changes sign, shifting its minimum from $a=0$ to $a=\pm\pi f_a$, implying that the nucleons source the axion, as first identified in \cite{Hook_2018,Balkin_2020}. For the standard QCD axion, this mechanism sets in only above a critical density of order twice the nuclear saturation density, a condition readily achieved in NS cores. If the stellar radius satisfies $R \sim {\cal O}(m_a^{-1})$, nucleons can source a nontrivial axion profile, with the field reaching values of order $\pi f_a$ at the core and gradually diminishing toward the surface, continuously matching onto the vacuum solution at large distances. Conversely, the backreaction of the axion condensate on the nucleons can be described as an effective shift in the nucleon mass, given by
\begin{equation}
m^*(a) \equiv m_N - \sigma_N \left[1 - g(a)\right],\label{eq:nucleon_effmass}
\end{equation}
with $m_N \approx 939~\mathrm{MeV}$ being the bare nucleon mass. For field values $a \sim \pi f_a$, this corresponds to a mass reduction of about $32~\mathrm{MeV}$
\cite{Gomez_2024}.

\subsection{Neutron star structure in the presence of axion condensates}
\label{Sect2B}

In order to derive the equilibrium NS structure equations, we assume spherical symmetry and adopt the usual interior Schwarzschild metric
\footnote{We use the metric signature $(+,-,-,-)$, which is standard in particle physics but opposite to the convention often employed in gravitation.}
in spherical coordinates $(r,\theta,\varphi)$,
\begin{equation}
ds^2 = \mathrm{e}^{2\Phi(r)}dt^2 - \mathrm{e}^{2\lambda(r)} dr^2 
- r^2 (d\theta^2 + \sin^2\theta d\varphi^2),
\label{eq:Schw}
\end{equation}
with $ \lambda(r) = - \frac{1}{2} \ln \left[1-\frac{2 GM(r)}{r}\right]$, where $M(r)$ is the gravitational mass enclosed within a sphere of radius $r$.

In the presence of the axion condensate, the static equilibrium of the system is described by the Tolman-Oppenheimer-Volkoff (TOV) equations, now including the contributions of the axion field, supplemented by the Klein-Gordon (KG) equation \cite{Springmann_2023, Balkin_2024},
\begin{widetext}
\begin{subequations}\label{eq:TOVaxionmat}
    \begin{align}
        p' &= -\left(\varepsilon + p\right)\left(\Phi'+\frac{n_s}{\varepsilon+p} \frac{\partial m^{*}}{\partial a}a'\right) ,\label{eq:TOVaxionmatp'} \\
        \Phi' &= \frac{1}{r^2}\Biggl(1 - \frac{2GM}{r}\Biggr)^{-1}\Biggl\{GM + 4\pi G r^3 \left[p - V(a) + \frac{(a')^2}{2} \left(1 - \frac{2GM}{r}\right)\right]\Biggr\}, \\
        M' &= 4\pi r^2 \left[\varepsilon + V(a) + \frac{(a')^2}{2} \left(1 - \frac{2GM}{r}\right)\right], \label{eq:TOVaxionmatM'} \\
        \intertext{and}
        a'' &\left(1 - \frac{2GM}{r}\right) + 2\frac{a'}{r} 
        \left[1 - \frac{GM}{r} - 2\pi G r^2 \left(\varepsilon - p + 2V(a)\right)\right] = \frac{\partial V}{\partial a} + n_s \frac{\partial m^*}{\partial a}.
    \end{align}
\end{subequations}
\end{widetext}
Here and henceforth, primes denote radial derivatives, whereas $p$ and $\varepsilon$ correspond to the pressure and energy density of matter, respectively.

Throughout the calculations, for simplicity, we will set $n_s = n_b$, where $n_b$ is the baryon density. The corrections proportional to the scalar density enter in the rhs of the TOV and the Klein--Gordon equations and this approximation may, at the perturbative level, induce small shifts in the mode frequencies. However, these shifts are comparable to, or smaller than, other uncertainties in the problem, such as those associated with the choice of EOS or the stellar mass. A more refined treatment of this term may become relevant in the future, if observational data reach the precision required to distinguish models differing at the $\approx 10 \%$ level in their predicted frequencies.

The system \eqref{eq:TOVaxionmat} is closed by an appropriate choice for the EOS, of the form $\varepsilon=\varepsilon(p, m^*(a))$. Note that the dependence of the EOS on the axion field, through the modified nucleon mass, is kept explicit. In principle, this implies that the chosen dense nuclear matter EOS would have to be rederived with $(a,\,f_a)$ as additional parameters (or, rather, the ratio $a/f_a$). In line with \cite{Gomez_2024}, we model NS matter using the BSk26 EOS \cite{BSk26}, with axion effects being reliably approximated by a reduction in the energy density, namely
\begin{equation}
    \varepsilon(n_b, a)\approx\varepsilon(n_b, 0)-n_b\sigma_N\left[1 - g(a)\right].\label{eq:eparticle_axion}
\end{equation}
Note that using the approximation \eqref{eq:eparticle_axion} and the first law of thermodynamics to derive $p(n_b, a)$ implies
\begin{equation}\label{eq:pressurefromeparticle_axion}
    p(n_b, a)=-\Biggl(\frac{\partial\left(\varepsilon/n_b\right)}{\partial\left(1/n_b\right)}\Biggr)_a\approx p\left(n_b, 0\right).
\end{equation}
In other words, to leading order, our approach involves incorporating the axion correction into the energy density through an effective mass term, while keeping the pressure--baryon density relation unchanged.

By numerically solving the coupled system of ordinary differential equations \eqref{eq:TOVaxionmat}, accompanied by the appropriate boundary conditions, one obtains the equilibrium structure of NSs in which the core is permeated by an axion field. Before proceeding with the computation of the oscillation modes, we first outline some of the general features of these solutions.

\begin{figure*}[htb!]
    \centering
    \includegraphics[width=\textwidth]{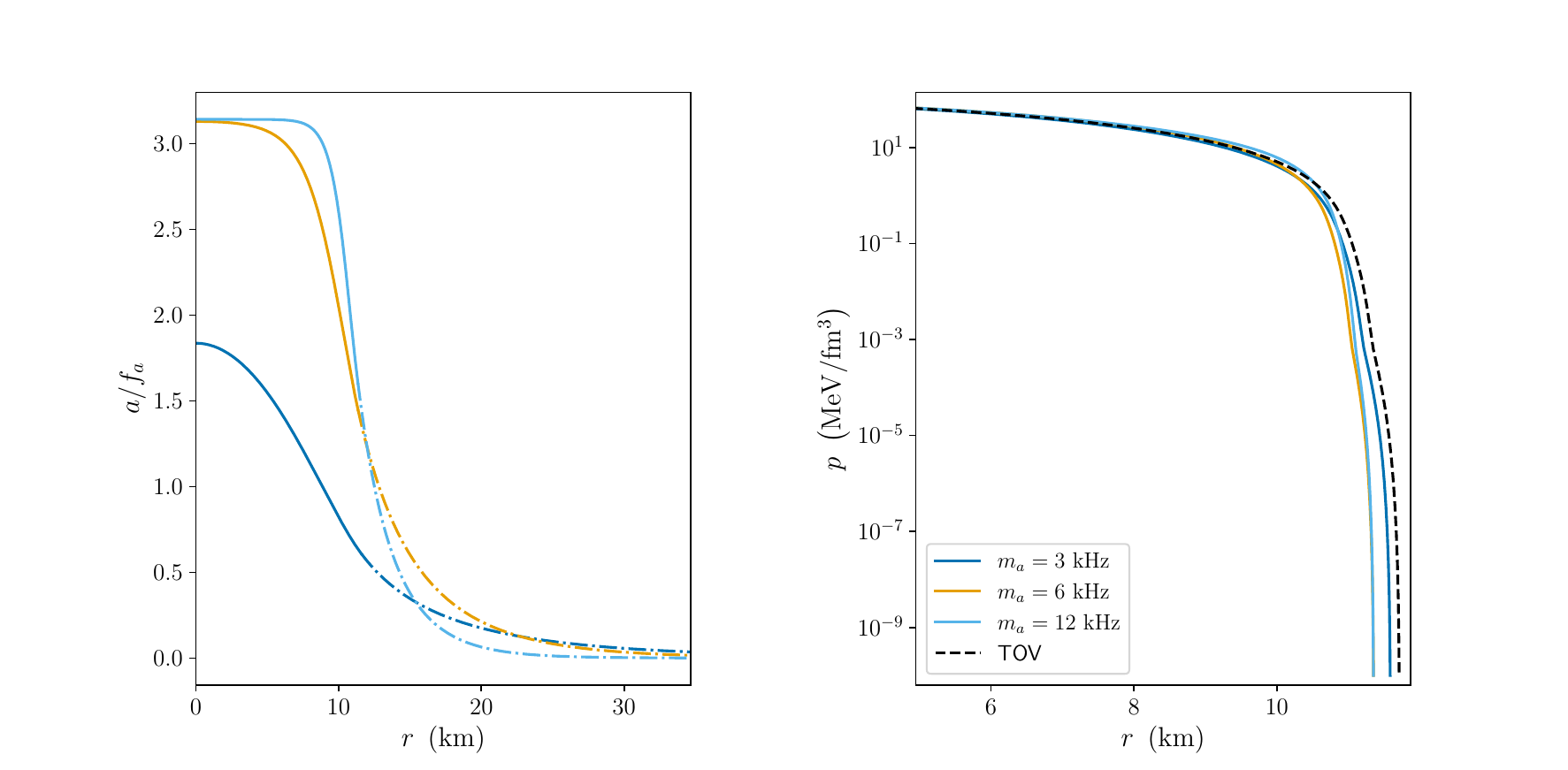}
    \caption{Normalized axion field $a/f_a$ (left) and pressure (right) profiles for different values of $m_a$ (in kHz) for $\epsilon=0.1$. The dash-dotted lines mark the axion profiles outside the NS. The central pressure of the NS is $p_0=100~\mathrm{MeV/fm^3}$ and its gravitational mass is approximately $1.5~M_{\odot}$ (with the latter exhibiting variations up to 2\% due to the corresponding variations in the energy content of the axion field profile for each $m_a$).}
    \label{fig:Profiles}
\end{figure*}

In Fig.~\ref{fig:Profiles} we show representative profiles of $a(r)$ and $p(r)$, for a NS with a central pressure of $p_0=100~\mathrm{MeV/fm^3}$.  We show results for three different values of $m_a$ (chosen to be 3, 6, and 12 kHz, corresponding to 12, 24, and 48 peV). 
For a typical radius of $R=12$ km, one finds $R/c \simeq 0.04$ ms, so that an axion mass of $m_a = 3$ kHz yields $m_a R \simeq 0.12$.  This value implies that the field varies slowly over length scales of the order of several stellar radii. 
On the other hand, for large axion masses ($m_a R \gg 1$), the axion amplitude remains nearly constant, close to $a \simeq \pi f_a$ (where the minimum of the potential is) in most of the NS interior, until it exhibits a sharp transition around the critical density, rapidly decaying to $a \simeq 0$ outside the star. 
This behavior is similar to a phase transition, when the NGS of nuclear matter (coexisting with a finite axion field) switches to normal nuclear matter (for which $a=0$). 
The details of the formation of the NGS depend on the control parameter $\epsilon$, with the induced changes in the structure of NSs or white dwarfs being used to place constraints on its value \cite{Balkin_2024,Balkin_2025}. In this work, we set $\epsilon=0.1$ for all our calculations, a value that lies within the still unconstrained region.

As $m_a$ decreases, the coupling to matter becomes less efficient, resulting in a smoother radial gradient of the field profile. For each choice of central pressure and $\epsilon$, one finds a critical axion mass below which the field is no longer sustained in the core, and the central value $a(0)$ is exponentially suppressed. Specifically, for $\epsilon=0.1$ and $p_0=100~\mathrm{MeV/fm^3}$, this occurs at $m_a \lesssim 2.8~\mathrm{kHz}$, below which no condensate occurs.

\begin{figure}
    \centering
    \includegraphics[width=0.5\textwidth]{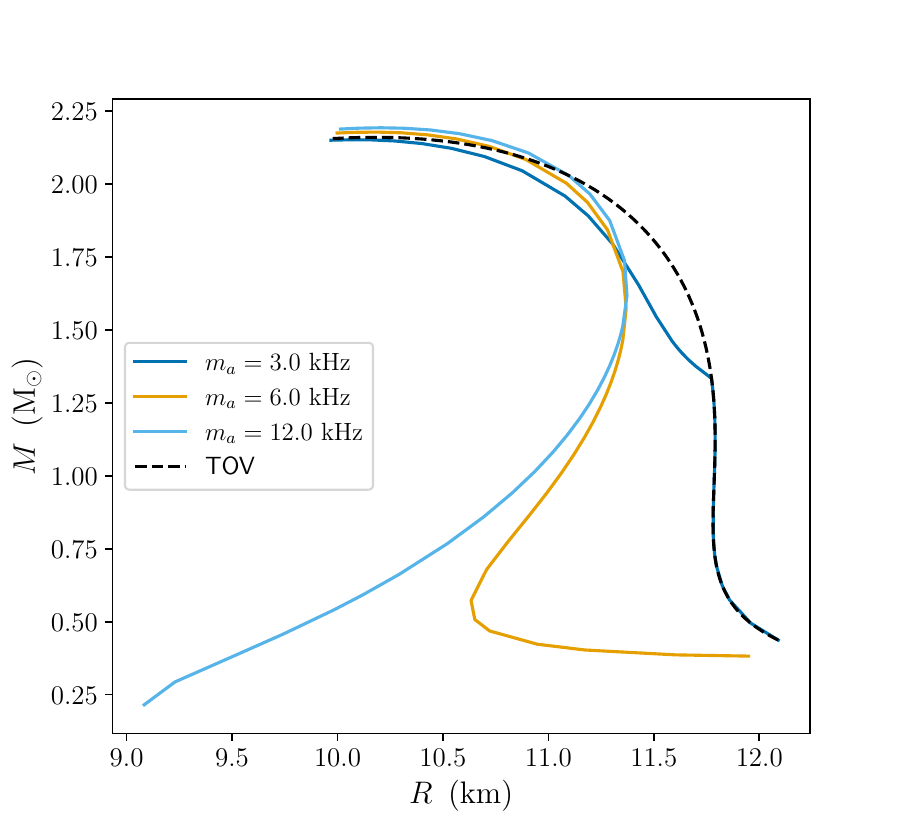}
    \caption{Mass–radius relation for NS models in the presence of an axion condensate, compared to the standard TOV solution for the same EOS (BSk26) without axions. The dashed black line represents the conventional TOV solutions without axions, whereas the solid colored curves illustrate the effect of including a bosonic condensate of axions inside the star, for different choices of the axion mass (3, 6, and 12 kHz) and for $\epsilon = 0.1$.}
    \label{fig:MRrelation}
\end{figure}

The way axion sourcing affects the NS structure may be seen more clearly in Fig.~\ref{fig:MRrelation}, where we compare the mass–radius ($M$--$R$) relations of different models, with and without the presence of an axion condensate. For light axions with $m_a = 3$ {kHz}, the $M$--$R$ relation is unaffected for lower masses (low central densities), but it is shifted toward smaller radii as the axion condensate grows in the center of the star, showing noticeable departures from the standard TOV curve already at moderate central densities. As the axion mass increases, the self-gravity of the condensate, combined with the effect of the gradient term ($\propto a'$) in Eq.~\eqref{eq:TOVaxionmatp'},  becomes more important and leads to more pronounced modifications in the stellar structure. This gradient term, playing a role similar to surface tension \cite{Gomez_2024}, is responsible for the sharp decrease of the axion field near the surface, an effect that compresses the NS outer layers, shrinking its envelope. This structural alteration leads to significant effects on the NS cooling behavior, thus placing constraints on the $\left(\epsilon, f_a\right)$ parameter space \cite{Gomez_2024, Kumamoto_2025}.

For $m_a = 12$ {kHz} the $M$--$R$ curve shape changes dramatically, resembling solutions of self-bound objects, such as strange stars, resulting in smaller radii (by about 1 km for a $1\,M_\odot$ NS). The formation of a \textit{bare} NS can be understood in terms of a NGS, reminiscent of strange quark matter, in the limit $f_a\rightarrow0$ \cite{Balkin_2024}. The critical value of $\epsilon$ for which the NGS forms depends on the EOS, albeit weakly. For realistic NS matter, previous studies have found that the NGS emerges when $\epsilon\lesssim0.1$ \citep{Gomez_2024,Kumamoto_2025}.

In general, the presence of an axion condensate tends to shrink the NS radius relative to the uncondensed case. Still, the effect on the maximum supported mass is very small because the global effect of the condensate is relatively smaller for stars with larger central densities, close to the maximum mass. Similar results were obtained in \cite{Balkin_2025}, where the authors examined the structure of NSs coupled to various light scalar fields, including the QCD axion. 
Their analysis differs from ours in that they adopted a free Fermi-gas EOS, which nevertheless captures most of the relevant phenomenology. Here, we demonstrate that the qualitative picture remains unchanged when a realistic EOS is used. These results highlight how accurate measurements of NS masses and radii could potentially serve as probes of light bosonic fields coupled to dense nuclear matter.

\section{Radial oscillations and quasinormal modes}\label{sec:radosc}

\subsection{Linear perturbation equations}

Following the standard procedure, we perturb the system of equations governing a spherically symmetric fluid coupled to a bosonic field around the background, as derived in Eq.~\eqref{eq:TOVaxionmat}. By introducing small perturbations to the metric $\delta\Phi, \delta\lambda$, the fluid thermodynamical variables $\delta p, \, \delta\varepsilon, \, \delta n_b$, and the axion field $\delta a$, we derive the linearized equations that capture the evolution of these perturbations, accounting for their mutual interactions and the coupling between the fluid and the axion field in the curved spacetime.

The components of the four-velocity perturbation are given by
\begin{equation}\label{eq:4vel_sphsymmetry2}
\delta u^\mu=\left(-\delta\Phi e^{-\Phi}, \, e^{-\Phi}\dot{\xi}^r, \, 0, \, 0\right),
\end{equation}
where we used the fact that $\delta\left(u^{\mu}u_{\mu}\right)=0$ and denoted the fluid radial displacement by $\xi^r$. The EOS and the first law of thermodynamics allow us to relate the perturbations of all thermodynamic quantities as
\begin{subequations}\label{eq:eos1stlaw_relations}
    \begin{align}
        \delta p & = c_s^2\left(\delta\varepsilon-n_b\frac{\partial m^*}{\partial a}\delta a\right),\\
        \delta p&=\frac{(\varepsilon+p)c_s^2}{n_b}\delta n_b,\label{eq:eos1stlaw_relations_deltanb}
    \end{align}
\end{subequations}
where $c_s^2\equiv\left(\partial p/\partial\varepsilon\right)_a$ is the square of the speed of sound and we assumed that $\left(\partial p/\partial a\right)_{n}\approx 0$ [see Eq.~\eqref{eq:pressurefromeparticle_axion}].

In principle, thermodynamic relations involving perturbed variables, such as those given by Eq.~\eqref{eq:eos1stlaw_relations}, strictly apply to Lagrangian perturbations \cite{Gittins_2025, Cox_1980}. As opposed to the Eulerian perturbations used here, which describe the change in a quantity $A$ at a fixed point in space, Lagrangian perturbations describe the respective change in a quantity, $A$, along the trajectory of a displaced fluid element, and are defined as $\Delta A = \delta A + \xi^rA'.$ However, since the same EOS is used for both the background and perturbed systems, it can be demonstrated that the relations in \eqref{eq:eos1stlaw_relations} hold for both Eulerian and Lagrangian perturbations. 

Building on the equilibrium background configurations introduced in Sec.~\ref{Sect2B}, we impose a harmonic time dependence $e^{-i\omega t}$ 
on all perturbations. To first order in the perturbation amplitudes, this leads to the following system of equations: 
\begin{widetext}
\begin{subequations}\label{eq:Einseq_sphsymmetry_pert}
\begin{align}
    &e^{-2\lambda}\delta a''+e^{-2\lambda}\delta a'\left(\Phi'-\lambda'+\frac{2}{r}\right)\nonumber\\
    &-2\delta\lambda\left(\frac{\partial V}{\partial a}+n_b\frac{\partial m^*}{\partial a}\right)+e^{-2\lambda} a'\left(\delta\Phi'-\delta\lambda'\right)\nonumber\\ &-\left(\frac{\partial^2 V}{\partial a^2}+n_b\frac{\partial^2 m^*}{\partial a^2}-\omega^2 e^{-2\Phi}\right)\delta a=\delta n_b\frac{\partial m^{*}}{\partial a},\label{eq:KGQCDaxmetric_sphsymmetry_pert}\\
    \delta\Phi' - \delta\lambda\left(\frac{1}{r} + 2\Phi'\right)&=4\pi G r e^{2\lambda}\left[\delta p-\frac{\partial V}{\partial a}\delta a - \delta\lambda e^{-2\lambda}(a')^2+e^{-2\lambda}a'(\delta a)'\right],\label{eq:Einseq_sphsymmetry_rrpert}\\
    \delta{\lambda}&=-4\pi Gre^{2\lambda}\left[(\varepsilon+p)\xi^r-e^{-2\lambda}a'\delta a\right]\label{eq:Einseq_sphsymmetry_rtpert},\\
    (\delta p)'+\left(\delta \varepsilon+\delta p\right)\Phi'&+\left(\varepsilon+p\right)\delta \Phi'\nonumber-\omega^2\left(\varepsilon+p\right)e^{2\left(\lambda-\Phi\right)}\xi^r=\nonumber\\
    &=-\frac{\partial m^*}{\partial a}\left(\delta n_b a'+ n_b(\delta a)'\right)-n_b\frac{\partial^2m^*}{\partial a^2}a'\delta a,\label{eq:fluideq_sphsymmetryCh16pert}\\
    \shortintertext{and}
    \left[\left(\varepsilon+p\right)\xi^r\right]'+&\left(\varepsilon+p\right)\xi^r\left(\frac{2}{r}+\lambda'+\Phi'\right)=-\Biggl(\frac{\delta p}{c_s^2}+\left(\varepsilon+p\right)\delta\lambda\Biggl).
\end{align}
\end{subequations}
\end{widetext}

\subsection{Boundary conditions}

Imposing regularity of all the perturbed variables at the origin implies the following expansions near the stellar center:
\begin{subequations}
    \begin{align}
    \delta a(r)&\sim\delta a(0)+\frac{\delta a''(0)}{2}r^2,\\
    \delta\Phi(r)&\sim\delta\Phi(0)+\frac{\delta \Phi''(0)}{2}r^2,\\
    \delta\lambda(r)&\sim\frac{\delta \lambda''(0)}{2}r^2,\\
    \shortintertext{and}
    \xi(r)&\sim\xi'(0)r.
    \end{align}
\end{subequations}
Furthermore, hydrostatic equilibrium at the stellar surface requires that the Lagrangian perturbation of the pressure $\Delta p$ vanishes, namely,
\begin{equation}\label{eq:DeltapRboundarycond}
    \Delta p(R)=0.
\end{equation}

The last boundary condition to be specified controls the behavior of $\delta a$ at large distances ($r\rightarrow\infty$). For an isolated, oscillating NS, we require that no incoming axion radiation be permitted. To impose this condition, we must examine the solutions of Eq.~\eqref{eq:Einseq_sphsymmetry_pert} when $r\rightarrow\infty$.

Outside the star, the fluid variables (both perturbations and background) vanish and the system \eqref{eq:Einseq_sphsymmetry_pert} is reduced to the following equations for the metric and axion variables:
\begin{widetext}
\begin{subequations}\label{eq:Einseq_sphsymmetry_pert_outside}
\begin{align}
    &e^{-2\lambda}\delta a''+e^{-2\lambda}\delta a'\left(\Phi'-\lambda'+\frac{2}{r}\right)\nonumber\\
    &-2\delta\lambda\left(\frac{\partial V}{\partial a}+n_b\frac{\partial m^*}{\partial a}\right)+e^{-2\lambda} a'\left(\delta\Phi'-\delta\lambda'\right)\nonumber\\ &-\left(\frac{\partial^2 V}{\partial a^2}-\omega^2 e^{-2\Phi}\right)\delta a=0,\label{eq:KGQCDaxmetric_sphsymmetry_pert_outside}\\
    \delta\Phi' - \delta\lambda\left(\frac{1}{r} + 2\Phi'\right)& =4\pi G r e^{2\lambda}\left[-\frac{\partial V}{\partial a}\delta a - \delta\lambda e^{-2\lambda}(a')^2+e^{-2\lambda}a'(\delta a)'\right] ,\label{eq:Einseq_sphsymmetry_rrpert_outside}\\
    \shortintertext{and}
    \delta\lambda &=4\pi G r a'\delta a\label{eq:Einseq_sphsymmetry_rtpert_outside}.
\end{align}
\end{subequations}
\end{widetext}
Since the amplitude of the axion field decays exponentially ($a \propto e^{-m_a r}/r$), terms proportional to $a$ or its derivatives become negligible at large distances. In addition, we know that the gravitational mass $M$ approaches a constant value, as the only contributions to $M'$ outside the NS arise from terms proportional to the exponentially vanishing axion field. With this, equations in~\eqref{eq:Einseq_sphsymmetry_pert_outside} are simply reduced to
\begin{equation}\label{eq:KGQCDaxmetric_sphsymmetry_pert_outside2}
     \left(1-\frac{2GM}{r}\right)\delta a'' + \frac{2}{r}\left(1-\frac{GM}{r}\right)\delta a'-\left(m_a^2-\frac{\omega ^2}{1-2GM/r}\right)\delta a = 0,
\end{equation}
recovering the KG equation for perturbations of a scalar field in a Schwarzschild background. The asymptotic behavior of $\delta a$ at infinity comprises two linearly independent solutions
of the form
\begin{equation}\label{eq:delta_ainf}
    \delta a\sim \frac{A}{r}e^{ikr}+\frac{B}{r}e^{-ikr},
\end{equation}
where $k\equiv\sqrt{\omega^2-m_a^2}$, and $A$ and $B$ are constants. The physical condition of no incoming axion radiation is thus equivalent to setting $B=0$, thereby selecting the outgoing solution.

Equations in~\eqref{eq:Einseq_sphsymmetry_pert}, together with the prescribed boundary conditions, define an eigenvalue problem describing the coupled oscillations of the NS and the axion field. The numerical procedure used to obtain the mode solutions, together with some important subtleties related to cases in which the imaginary part of $k$ in Eq.~\eqref{eq:delta_ainf} is large, are addressed in Appendix~\ref{app:numerical_procedure}.

Since the axion perturbation $\delta a$ satisfies a wave equation and can propagate beyond the star's surface, part of the NS oscillation energy can be transferred to the axion field, resulting in the emission of axions to infinity and a corresponding damping of the stellar oscillations. This mechanism is analogous to the emission of gravitational radiation from quadrupolar oscillations in NSs \cite{Thorne_1967, Lindblom_1983}. However, unlike GWs, which require a nonzero quadrupole moment and are only sourced by modes with a multipole degree $l\geq 2$, the scalar nature of the axion field permits axionic radiation even from purely radial pulsations $(l=0)$. This problem is similar to others discussed in the literature involving fluids coupled to scalar degrees of freedom, for example in the context of scalar-tensor theories of gravity \cite{Sotani_2004, Sotani_2005, Krueger_2021, Mendes_2018}. What makes this case distinctive is the fact that it probes a direct coupling between the scalar field and the fluid, as opposed to an indirect coupling through gravity.

\section{A simplified description of the axion-fluid coupling} \label{sec:simple}

Before proceeding, let us examine a simplified version of Eq.~\eqref{eq:Einseq_sphsymmetry_pert} that retains the essential physics and aids in interpreting the results for the full system. To reduce complexity, we neglect metric perturbations, background derivatives, assume planar symmetry, and utilize the following simplifications:
\begin{subequations}
    \begin{align}
        \frac{\partial m^*}{\partial a}&\approx -\frac{\beta\sigma_N}{4f_a},\\ \frac{\partial^2 V}{\partial a^2}+n_s\frac{\partial^2 m^*}{\partial a^2}&=\left(m_a^*\right)^2\approx m_a^2\left(1-\frac{\sigma_Nn_b}{\epsilon m_{\pi}^2f_{\pi}^2}\right).
    \end{align}
\end{subequations}
 We then arrive at the following system of equations:
\begin{subequations}\label{eq:toymodeleqs}
\begin{align}
    &\delta p'' + \frac{\omega^2}{c_s^2}\delta p =\frac{\beta\epsilon m_{\pi}^2f^2_{\pi}}{4}\left(\frac{n_b}{n_c}\right)\delta \theta'',
    \\
    &\delta \theta'' + \left(\omega^2 - m_a^2\left(1-\frac{n_b}{n_c}\right) \right)\delta \theta  =
    -m_a^2\left(\frac{n_b}{n_c}\right)\left(\frac{\delta p}{\varepsilon +p}\right)\frac{1}{c_s^2},
\end{align}
\end{subequations}
where we used the relation \eqref{eq:eos1stlaw_relations_deltanb} and defined 
$\delta\theta\equiv\frac{\delta a}{f_a}$.

To cast the equations into a more compact, dimensionless form involving only a few constant parameters, we introduce the following notation and approximations. 
From this point on, all spatial derivatives are taken with respect to the dimensionless variable $\bar{x} = m_a x$, and frequencies are expressed in units of $m_a$, i.e., $\bar{\omega} = \omega/m_a$. In addition, we approximate the radial density profile by a step function,
\begin{equation}
n_b(x) =
    \begin{cases}
        \alpha n_c & \text{if } \bar{x} \leq \bar{x}_0 \\
         0 & \text{if } \bar{x} > \bar{x}_0
    \end{cases}
\end{equation}
with $\alpha$ being a constant. Consistently, the axion background is assumed to be a step function with $\theta = \pi$ if $\alpha > 1$ and $\theta=0$ otherwise. The distance $\bar{x}_0$ represents the NS radius in units of $m_a$. Note that the axion effective mass at finite density
$(m_a^{*})^2 = m_a^2 {\left(1-\alpha\right)}$ can be negative if $\alpha > 1$.

For $\bar{x}> \bar{x}_0$, the fluid perturbations vanish, and we are left with the axion equation
\begin{subequations}
\begin{align}
\delta \theta'' + \left(\bar{\omega}^2 - 1 \right)\delta \theta  = 0.
\end{align}
\end{subequations}

For $\bar{x}<\bar{x}_0$, 
we also define the dimensionless variable $\delta \tilde{p} \equiv \frac{4}{\beta\epsilon m_{\pi}^2f^2_{\pi}}  \delta p$, 
and the dimensionless coupling parameter 
$$
\lambda^2 \equiv 
\frac{m_a^2 f_a^2 }{(\varepsilon +p) c_s^2}.
$$
Since $m_a^2f_a^2 \sim \epsilon ~ 13 ~\mathrm{MeV/fm^3}$
and $c_s^2 \approx 0.3$, we have for 
$(\varepsilon+p)\sim 500~\mathrm{MeV/fm^3}$ that 
$\lambda^2\sim \epsilon 0.02$.

Thus, we can write
\begin{subequations}
\begin{align}
    &\delta \tilde{p}'' + 
    \frac{\bar{\omega}^2}{c_s^2} \delta \tilde{p} =
    - \lambda^{2} \alpha^2 \delta\tilde{ p} 
    - \alpha \left(\bar{\omega}^2 - 1 + \alpha \right) \delta \theta ,
    \\
    &\delta \theta'' + \left(\bar{\omega}^2 - 1 +\alpha \right)\delta \theta  =
    - \lambda^{2} \alpha {\delta\tilde{p}},
\end{align}
\end{subequations}
or in matrix form
\begin{eqnarray}
\begin{pmatrix}
    \delta\tilde{p}\\
    \delta \theta
\end{pmatrix}''&= -
\begin{pmatrix}\label{eq:toymodel_matrix}
    \frac{\bar{\omega}^2}{c_s^2}+ \alpha^2\lambda^2 \; \; & \alpha(\bar{\omega}^2 - 1 +\alpha) \\
    \alpha \lambda^2 \; \; & \bar{\omega}^2 - 1 +\alpha
\end{pmatrix}
\begin{pmatrix}
    \delta\tilde{p}\\
    \delta \theta
\end{pmatrix} \equiv -\mathbb{M}
\begin{pmatrix}
    \delta\tilde{p}\\
    \delta \theta
\end{pmatrix}
\end{eqnarray}

The eigenvalues $\bar{k}^2$ of the matrix $\mathbb{M}$ obey the
characteristic equation
\begin{equation}\label{eq:detMeq}
    \left( \bar{\omega}^2 -1 + \alpha-\bar{k}^2\right) \left(\bar{\omega}^2-\bar{k}^2 c_s^2\right) 
    - \alpha^2 \lambda^2 \bar{k}^2 {c_s^2} =0
\end{equation}
which leads to a quadratic equation in $\bar{\omega}^2$,
\begin{equation}
\bar{\omega}^4 - \bar{\omega}^2 
    \left( \bar{k}^2(1+c_s^2) + (1-\alpha) \right) 
    + c_s^2 \bar{k}^2 \left( \bar{k}^2 + (1-\alpha) - \alpha^2 \lambda^2 \right) =0
\end{equation}
with solutions
$$
\bar{\omega}^2 = \frac{ \bar{k}^2 + 1-\alpha + \bar{k}^2c_s^2
\;\pm\;\sqrt{\Big( \bar{k}^2 + 1-\alpha - \bar{k}^2c_s^2 \Big)^2
+ 4 c_s^2 \bar{k}^2 \alpha^2 \lambda^2}}{2}.
$$

\paragraph*{Case $\lambda\rightarrow0$.}
Consider first the decoupling limit $\lambda=0$.
In this situation, there are two possible independent classes of modes in the interior: discrete, confined fluid-dominated modes with a dispersion relation $\bar{\omega}^2 = \bar{k}^2c_s^2$, and the spectrum of axion modes with dispersion relation $\bar{\omega}^2 = \bar{k}^2 +1-\alpha$. 

In the exterior $\bar{\omega}^2 = \bar{k}^2 +1$, so that:
\begin{itemize}
\item For $\bar{\omega}^{2} < 1$, the axion field becomes evanescent, decaying as $\sim e^{-\kappa \bar{x}}$. 
\item For $\bar{\omega}^{2} > 1$, the axion field admits plane-wave solutions $e^{\pm i \bar{k}_{out} x}$ with $\bar{k}_{out}=\sqrt{\bar{\omega}^{2}-1}$.
\end{itemize}

However, the restrictions imposed by the boundary conditions and the continuity of the function and its derivative are worth discussing. The interior and exterior solutions satisfying the corresponding boundary conditions are
\begin{align}
    &\delta \theta=A\left(e^{i\bar{k}_{a}\bar{x}}+e^{-i\bar{k}_{a}\bar{x}}\right) & \quad \bar{x}<\bar{x}_0,\\
    &\delta\theta= C e^{i\bar{k}_{out}\bar{x}} & \quad \bar{x}>\bar{x}_0,
\end{align}
where 
$\bar{k}_{a}^2\equiv \bar{\omega}^2-1+\alpha$, and
we assume $\mathrm{Re}(\bar{\omega}) \geq 0$.
Imposing continuity of $\delta\theta$ and $\delta\theta'$ at $x_0$ leads to the condition
\begin{equation}\label{eq:trasc_eq_toymodel0}
    \bar{k}_a \tan\left(\bar{k}_{a}\bar{x}_0\right)=-i \bar{k}_{out}.
\end{equation}
If we look for solutions with $\bar\omega^2\in\mathbb{R}$, the above equation only admits solutions when 
$1 -\alpha<\bar{\omega}^2<1$.
Thus, the radial dependence of $\left(m_a^*\right)^2$ comes into play. Inside a realistic NS, the effective mass depends on $n_b$ and $\theta$, resulting in a nontrivial potential for $\delta\theta$, which may allow the existence of axion quasinormal modes (QNMs), or even quasibound axion modes [for cases in which $\omega^{2} > (m_a^*)^{2}$ inside but $\omega^{2} < m_a^{2}$ outside]. Such modes would be localized in the interior while slowly leaking out via tunneling, and would appear in the low-frequency axion spectrum.

\paragraph{Case $\lambda \neq 0$.}
Let us now turn to the more general coupled case with $\lambda \neq 0$.
The first natural expectation is the hybridization of modes:
discrete fluid oscillations couple to the axion spectrum and are promoted to QNMs with complex frequencies, where the imaginary part encodes damping through axion radiation.
The axion spectrum remains present, but may display resonant features in the vicinity of the fluid QNM frequencies.

To proceed with the discussion, it is convenient to invert $\bar{\omega}^2(\bar{k})$ and define $\bar{k}_s^2(\bar{\omega})\equiv\frac{\bar{\omega}^2}{c_s^2}$, $\bar{k}_a^2(\bar{\omega}) \equiv \bar{\omega}^2-1+\alpha$, so that we can now write explicitly
\begin{subequations}
\begin{align}
    \bar{k}_{\pm}^2 & =\frac{ \bar{k}_a^2+\bar{k}_s^2+\alpha^2\lambda^2\;\pm\;\sqrt{\Delta(\bar{\omega})}}{2 },\label{eq:kplusminus}\\
\Delta(\bar{\omega}) &=\left(\bar{k}_a^2-\bar{k}_s^2+\alpha^2\lambda^2\right)^2+4\alpha^2\lambda^2\bar{k}_s^2.
\end{align}
\end{subequations}
It can be shown that, for $\bar{\omega}^2\in\mathbb{R}$,
$\bar{k}^2_+ > \max{(\bar{k}_a^2, \bar{k}_s^2)} > \min{(\bar{k}_a^2, \bar{k}_s^2)}
> \bar{k}^2_- > 0.$ 

By diagonalizing the system \eqref{eq:toymodel_matrix}, one may introduce characteristic variables, 
which admit plane-wave expansions with wave numbers
$\bar{k}_{\pm}$. Enforcing the relevant boundary conditions and matching the interior solution to purely outgoing waves in the exterior leads to the following relation: 
\begin{equation}\label{eq:trasceq_deltapprime0}
    f(\bar{\omega}) \equiv (\bar{k}_a^2-\bar{k}_-^2)  \bar{k}_+  \tan{(\bar{k}_+\bar{x}_0)} -  (\bar{k}_a^2-\bar{k}_+^2) \bar{k}_- \tan{(\bar{k}_-\bar{x}_0)} + i \bar{k}_{out} \sqrt{\Delta} = 0,
\end{equation}
whose solutions are the QNMs. 
It can be readily shown that in the limit $\lambda = 0$, the above equation simplifies to Eq.~\eqref{eq:trasc_eq_toymodel0}.

\begin{figure}
    \centering
    \includegraphics[width=0.9\textwidth]{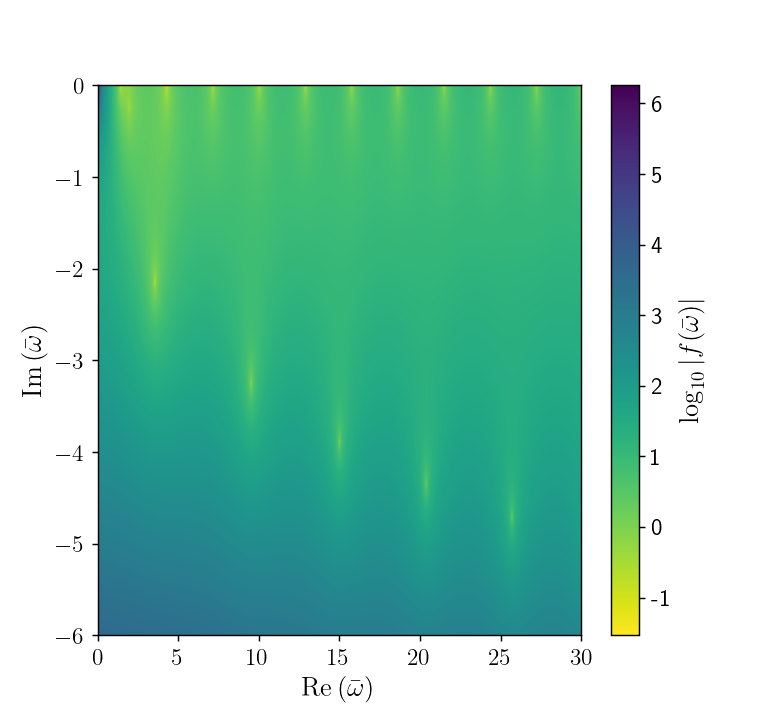}
    \caption{QNM spectrum landscape for the simplified model. The plot displays the values of Eq.~\eqref{eq:trasceq_deltapprime0} in the complex plane, for $\lambda^2=0.001, \alpha=10, c_s^2=0.3,$ and $x_0=0.6$. The bright yellow regions indicate the locations of the QNMs, with the two distinct families clearly identifiable. }
    \label{fig:toymodelfunction_deltapprime0=0}
\end{figure}

Figure~\ref{fig:toymodelfunction_deltapprime0=0} shows the landscape of QNMs in the complex frequency plane, for typical values of the parameters in a NS.
The horizontal axis represents the real part of the frequency, $\mathrm{Re}(\bar{\omega})$, and the vertical axis represents the imaginary part, $\mathrm{Im}(\bar{\omega})$. 
The color scale indicates the magnitude of the function, $\log_{10}|f(\bar{\omega})|$, such that the zeros of the function correspond to localized minima (bright yellow spots) in the plot. 
Two distinct families of modes are clearly visible: close to the real axis one finds the \emph{fluid-dominated modes}, which correspond to oscillatory solutions with relatively low damping, as seen from the clustering of minima near the real axis. 
In contrast, at larger negative imaginary parts, a second family appears (the \emph{axion modes}), which are strongly damped, as indicated by their deeper positions in the complex plane. 
This clear separation between weakly damped fluid-dominated modes and highly damped axion modes highlights the characteristic structure of the QNM spectrum.

\section{Radial modes in a realistic neutron star with an axion condensate}\label{sec:res}

We will now proceed with the computation of the radial mode spectrum of a NS, described by the BSk26 EOS and coupled to an axion field, by solving the eigenvalue problem outlined in Sec.~\ref{sec:radosc}. Our background model is an $M=1.53~M_\odot$, $R=11.7~\mathrm{km}$ NS, with a central pressure of $p_0=100~\mathrm{MeV/fm^3}$.

In order to establish a baseline for the mode behavior, we first solve the problem in the absence of the axion field.  The frequencies of the first four radial modes for this case are given in Table~\ref{tab:modescanp0100eps0.1radialmodes}. Subsequently, to investigate the variation of the mode frequencies with respect to $m_a$, we will span the range where the axion field is dynamically sourced and the axion mass is of the order of kHz---specifically, we will consider axion masses lying in the range $[2.8,12]$ kHz.

The axion condensate influences the mode frequencies for two different reasons: (i) by modifying the star’s background structure and (ii) by altering the perturbation equations governing the oscillations. To disentangle the physical origin of the induced frequency shifts, the perturbation equations in \eqref{eq:Einseq_sphsymmetry_pert} are solved both with and without the perturbations of the axion field (while always retaining the effects of the axion on the background structure).

\begin{figure*}[t]
    \centering
        \includegraphics[width=0.4\textwidth]{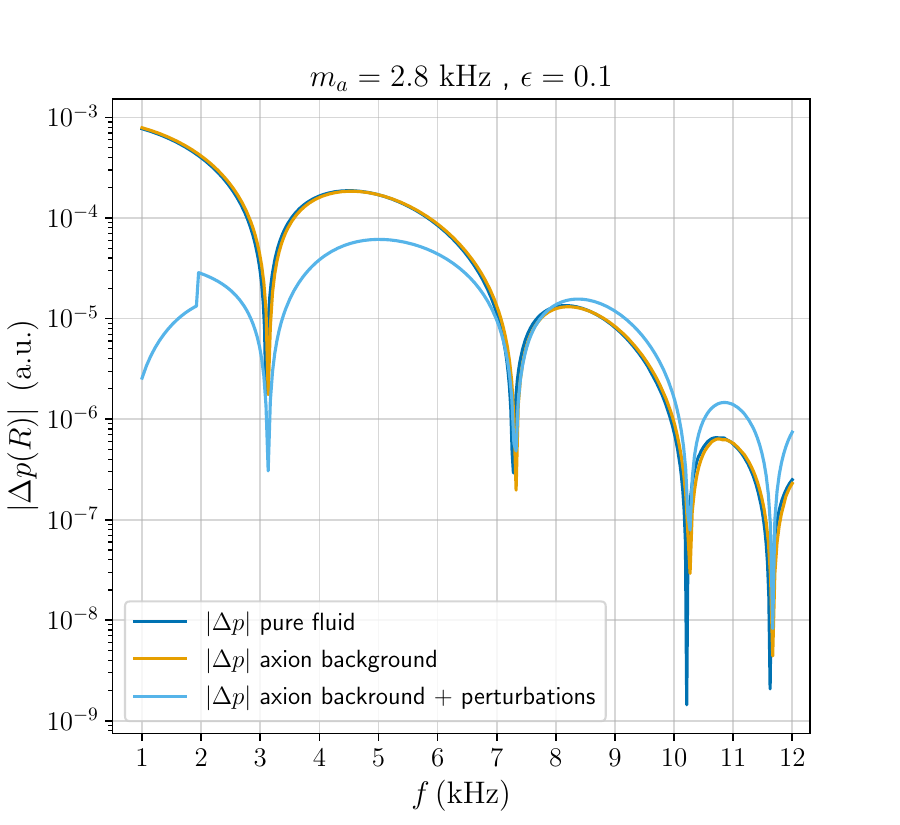} 
        \includegraphics[width=0.4\textwidth]{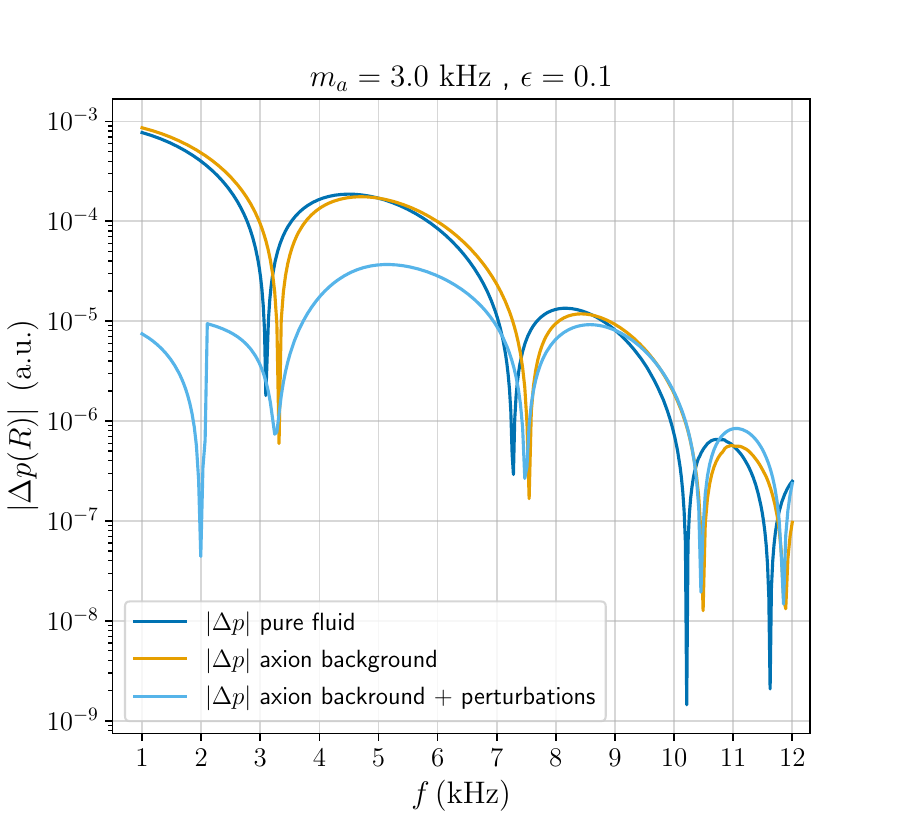} \\
        \includegraphics[width=0.4\textwidth]{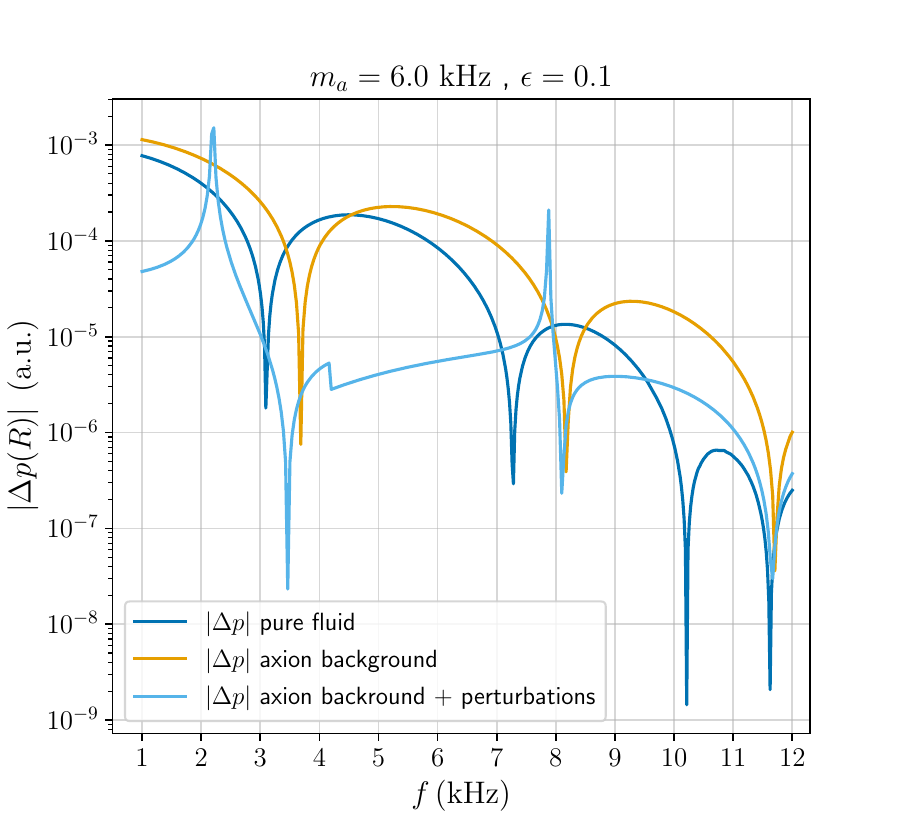} 
        \includegraphics[width=0.4\textwidth]{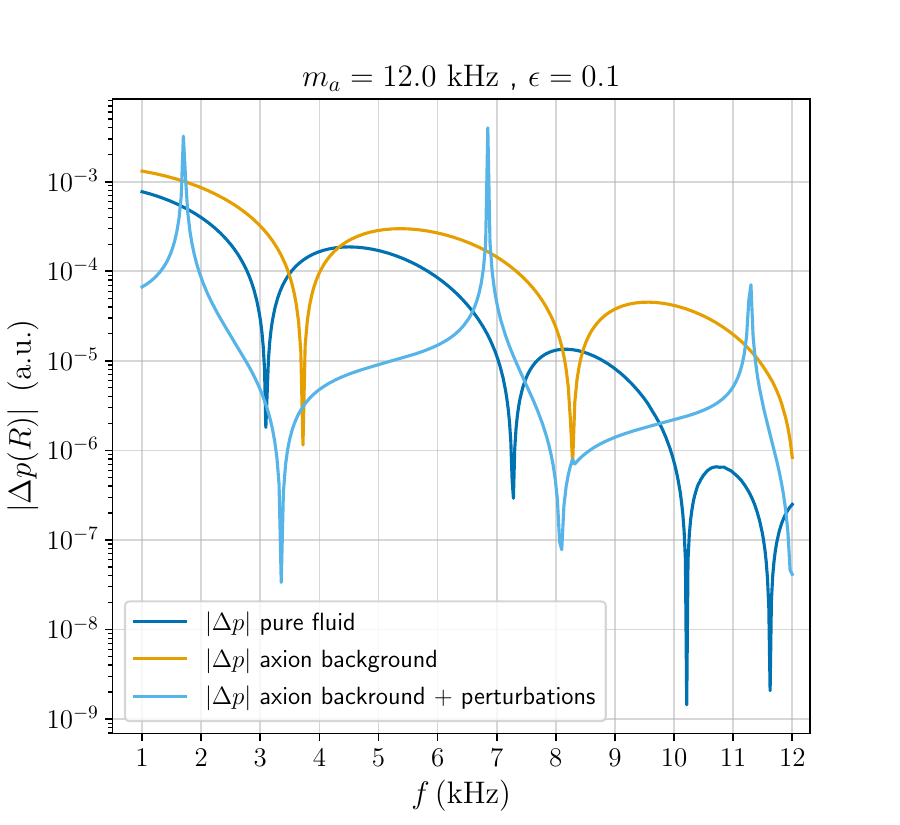} \\
    \caption{$\left|\Delta p(R)\right|$ (in arbitrary units) as a function of the real part of the frequency $f$ (in kHz), for $\epsilon=0.1$ and different axion masses $m_a$ (in kHz). The three curves correspond to the solution without an axion field (blue), the solution for which the axion perturbations are omitted (orange), and the full solution (cyan).}
    \label{fig:fourfigsmodescanp0100eps0.1}
\end{figure*}

Our method for identifying the fluid-dominated modes is illustrated in Fig.~\ref{fig:fourfigsmodescanp0100eps0.1}, where we plot the absolute value of the Lagrangian perturbation of the pressure at the NS surface, $|\Delta p(R)|$, as a function of mode frequency for four selected parameter combinations. The three curves correspond to the solution without an axion field (blue), the solution for which the axion perturbations are omitted (orange), and the full solution (cyan). The locations of the mode frequencies can be clearly recognized by the deep, pronounced minima, where $|\Delta p(R)|$ approaches zero.

\begin{table}[ht]
    \centering
    \begin{tabular}{d{2.2}d{2.2}d{2.2}d{2.2}d{2.2}}
        \hline\hline
        \multicolumn{1}{c}{$m_a~\mathrm{(kHz)}$} & \multicolumn{1}{c}{$f_0~\mathrm{(kHz)}$} & \multicolumn{1}{c}{$f_1~\mathrm{(kHz)}$} & \multicolumn{1}{c}{$f_2~\mathrm{(kHz)}$} & \multicolumn{1}{c}{$f_3~\mathrm{(kHz)}$} \\
        \hline
        \multicolumn{1}{c}{no axion} & 3.10 & 7.27 & 10.22 & 11.63 \\
        2.8      & 3.14 & 7.32 & 10.26 & 11.67 \\
        3.0      & 3.32 & 7.54 & 10.46 & 11.86 \\
        4.0      & 3.63 & 8.03 & 11.25 & 12.84 \\
        5.0      & 3.67 & 8.14 & 11.56 & 13.45 \\
        6.0      & 3.68 & 8.18 & 11.71 & 13.87 \\
        8.0      & 3.69 & 8.22 & 11.88 & 14.45 \\
        12.0     & 3.72 & 8.27 & 12.07 & 15.16 \\
        \hline\hline
    \end{tabular}
    \caption{Frequencies $f_n$ (in kHz) of the first four radial oscillation modes ($n$ denoting the mode overtone), for $\epsilon=0.1$ and different axion masses $m_a$ (in kHz), neglecting the axion perturbations [corresponding to the minima of $\left|\Delta p(R)\right|$ in the orange curves of Fig.~\ref{fig:fourfigsmodescanp0100eps0.1}]. The frequencies of the pure fluid modes, in the absence of an axion condensate, are also given for reference [corresponding to the minima of $\left|\Delta p(R)\right|$ in the blue curves of Fig.~\ref{fig:fourfigsmodescanp0100eps0.1}].}
    \label{tab:modescanp0100eps0.1radialmodes}
\end{table}

In Table~\ref{tab:modescanp0100eps0.1radialmodes} we summarize our results for the case where axion perturbations are neglected, using the same parameter combinations as in Fig.~\ref{fig:fourfigsmodescanp0100eps0.1}. As $m_a$ increases, the upwards shift in the oscillation frequencies of all fluid modes is, in this case, only attributed to the structural changes in the background NS induced by the axion condensate. The effect is more pronounced as the axion mass and the mode overtone increase. 

\begin{figure}
    \centering
    \begin{tabular}{ccc}
		\includegraphics[width=.9\textwidth]{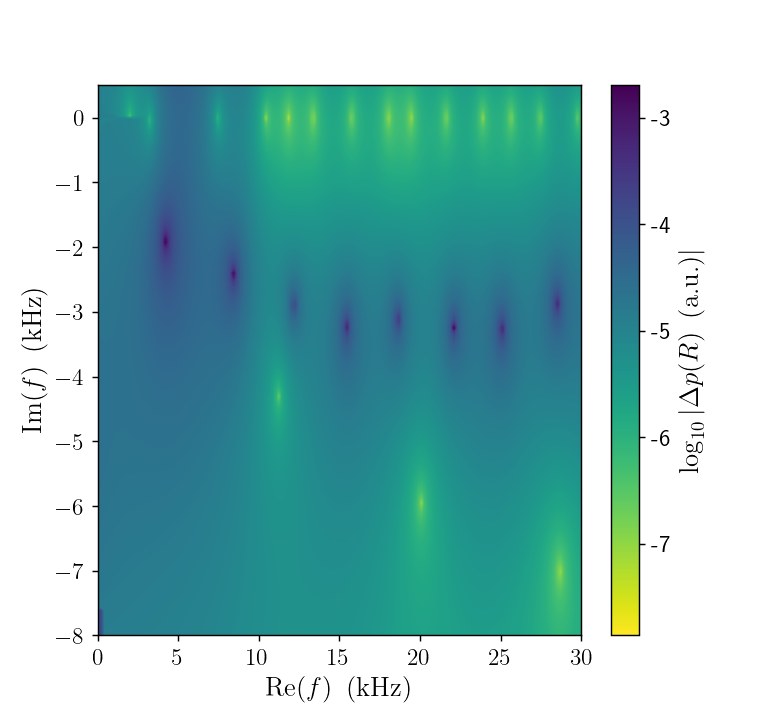} \\
	\end{tabular}
    \caption{QNM spectrum landscape for $m_a$ set to $3.0~\mathrm{kHz}$. It reveals the two families of modes: fluid-dominated ones with mild damping, and axion-led modes with stronger damping.
    The simplified model introduced in Sec.\ref{sec:simple} and illustrated in Fig.~\ref{fig:toymodelfunction_deltapprime0=0} showed the same qualitative characteristics as this plot.}
    \label{fig:p0100ma3kHzQNMspectrum}
\end{figure}

When axion perturbations are also taken into account, the real parts of the mode frequencies experience an additional shift. This shift partially compensates for the effect of the axion background, leading to a slight reduction in the mode frequencies compared to the case where the axion perturbations are neglected. A second, and arguably more significant, consequence is the appearance of an entirely new family of modes, tied to the oscillations of the axion field. A general view of the landscape of the QNMs is presented in Fig.~\ref{fig:p0100ma3kHzQNMspectrum}.
The insights from the simplified fluid-axion model in \mbox{Sec.~\ref{sec:simple}} provide valuable intuition for understanding the general case, which exhibits strong qualitative similarities.

\begin{figure}[htb!]
    \centering
    \includegraphics[width=0.9\textwidth]{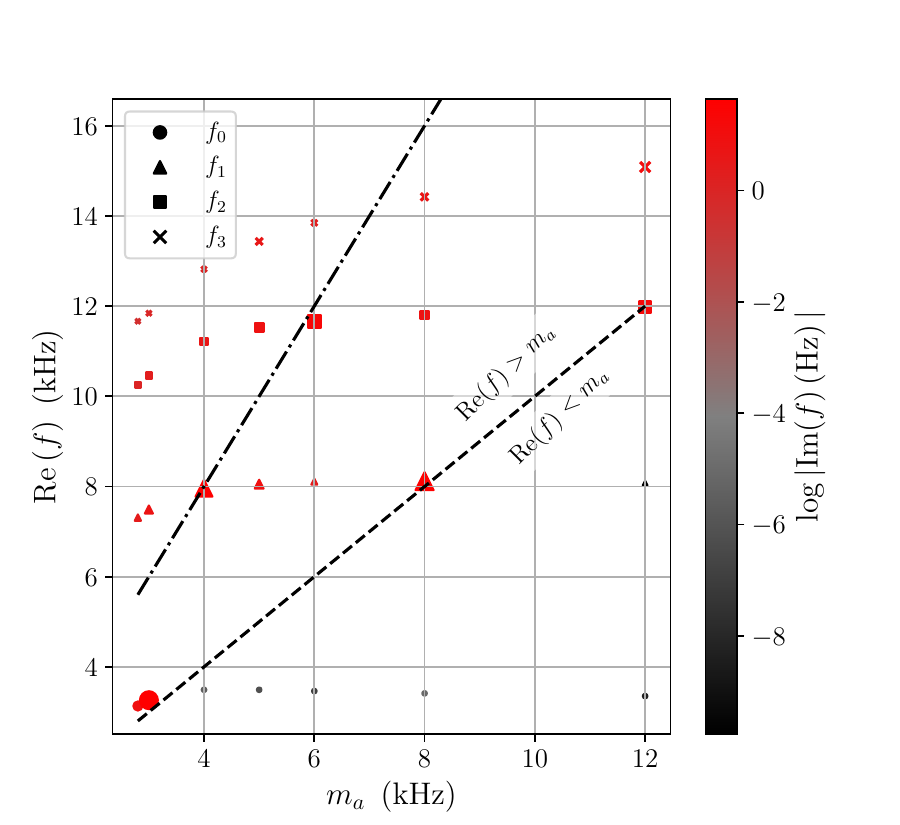}
    \caption{Real part of the mode frequency (kHz) against $m_a$ (kHz) for the fluid-dominated modes, as detailed in Table~\ref{tab:modescanp0100eps0.1radialmodesfluid}. The color scale shows $\log\left|\mathrm{Im}f\,(\mathrm{Hz})\right|$ and the marker size also grows proportionally to $\left|\mathrm{Im}f\right|$. We overplot with dashed and dash-dotted the two lines  
    corresponding to $\mathrm{Re}(f)=m_a$ and $\mathrm{Re}(f)=2m_a$, respectively.}
    \label{fig:RefTableIIgamma2}
\end{figure}

The precise oscillation frequencies of the fluid-dominated and the axion modes, including their imaginary parts, obtained by solving the full problem, are presented in Tables~\ref{tab:modescanp0100eps0.1radialmodesfluid} and \ref{tab:modescanp0100eps0.1radialmodesaxion}, respectively. An instructive illustration of the frequencies of the fluid-dominated modes (Table~\ref{tab:modescanp0100eps0.1radialmodesfluid}) is also presented in Fig.~\ref{fig:RefTableIIgamma2}. Our results show that, through their coupling with the new axion-led oscillations, the original fluid modes acquire a small imaginary component, corresponding to a relatively short damping time (on the order of seconds), significantly faster than other potential damping mechanisms, such as viscosity \citep{Cutler1990, Gusakov2013}. 

All qualitative features are clearly visible in Fig.~\ref{fig:RefTableIIgamma2}. Modes with $\mathrm{Re}(f) < m_a$ remain undamped (grey), whereas those above this threshold have damping times on the order of seconds. Notably, the modes located near $\mathrm{Re}(f) = m_a$ and $\mathrm{Re}f = 2m_a$ display the strongest damping, which can be understood as resonances with the natural frequencies of the emitted axion modes.

\begin{table}[ht]
    \centering
    \begin{tabular}{d{2.2}d{2.2}d{2.2}d{2.2}d{2.2}d{2.2}d{2.2}d{2.2}d{2.2}}
        \hline\hline
        \multicolumn{1}{c}{\multirow{2}{*}{$m_a~\mathrm{(kHz)}$}} & \multicolumn{2}{c}{$f_0$} & \multicolumn{2}{c}{$f_1$} & \multicolumn{2}{c}{$f_2$} & \multicolumn{2}{c}{$f_3$} \\
        \cline{2-9}
        & \multicolumn{1}{c}{Re$f$~(kHz)} & \multicolumn{1}{c}{Im$f$~(Hz)} & \multicolumn{1}{c}{Re$f$~(kHz)} & \multicolumn{1}{c}{Im$f$~(Hz)} & \multicolumn{1}{c}{Re$f$~(kHz)} & \multicolumn{1}{c}{Im$f$~(Hz)} & \multicolumn{1}{c}{Re$f$~(kHz)} & \multicolumn{1}{c}{Im$f$~(Hz)} \\
        \hline
        \multicolumn{1}{c}{no axion} & 3.10 & & 7.27 & & 10.22 & & 11.63 & \\
        2.8  & 3.13 & -8.80 & 7.31 & -2.60 & 10.25 & -1.04 & 11.66 & -0.394 \\
        3.0  & 3.26 & -43.7 & 7.49 & -6.10 & 10.46 & -2.15 & 11.84 & -0.854 \\
        4.0  & 3.49 & 0 & 7.96 & -38.9 & 11.21 & -4.55 & 12.82 & -1.06 \\
        5.0  & 3.50 & 0 & 8.05 & -8.38 & 11.52 & -7.74 & 13.44 & -4.02 \\
        6.0  & 3.47 & 0 & 8.11 & -2.05 & 11.66 & -22.5 & 13.85 & -2.01 \\
        8.0 & 3.42 & 0 & 8.12 & -36.2 & 11.80 & -6.71 & 14.42 & -5.22 \\ 
        12.0 & 3.36 & 0 & 8.09 &  0    & 11.98 & -12 & 15.08 & -11.4 \\
        \hline\hline
    \end{tabular}
    \caption{Frequencies $f_n$ (real part in kHz, imaginary part in Hz) of the first four radial oscillation modes ($n$ denoting the mode overtone), for $\epsilon=0.1$ and different axion masses $m_a$ (in kHz), based on solving the full problem [with the real parts corresponding to the minima of $\left|\Delta p(R)\right|$ in the cyan curves of Fig.~\ref{fig:fourfigsmodescanp0100eps0.1}]. The frequencies of the pure fluid modes, in the absence of an axion condensate, are also given for reference [corresponding to the minima of $\left|\Delta p(R)\right|$ in the blue curves of Fig.~\ref{fig:fourfigsmodescanp0100eps0.1}].}
    \label{tab:modescanp0100eps0.1radialmodesfluid}
\end{table}

\begin{table}[ht]
    \begin{tabular}{d{2.2}d{2.2}d{2.2}d{2.2}d{2.2}d{2.2}d{2.2}}
        \hline\hline
        \multicolumn{1}{c}{\multirow{2}{*}{$m_a~\mathrm{(kHz)}$}} & \multicolumn{2}{c}{$f^\mathrm{a}_0$} & \multicolumn{2}{c}{$f^\mathrm{a}_1$} & \multicolumn{2}{c}{$f^\mathrm{a}_2$} \\
        \cline{2-7}
        & \multicolumn{1}{c}{Re$f$~(kHz)} & \multicolumn{1}{c}{Im$f$~(kHz)} & \multicolumn{1}{c}{Re$f$~(kHz)} & \multicolumn{1}{c}{Im$f$~(kHz)} & \multicolumn{1}{c}{Re$f$~(kHz)} & \multicolumn{1}{c}{Im$f$~(kHz)} \\
        \hline
        2.8  & 0.749 &  0    & 10.50 & -4.30 & 19.29 & -5.82 \\
        3.0  & 2.00  &  0    & 11.23 & -4.31 & 20.06 & -5.94 \\
        4.0  & 2 & -1.4 & 16.36 & -3.45 & 22.99 & -5.37 \\
        5.0  & 1.8  & -2 & 20.30 & -2.22 & 25.54 & -4.70 \\
        6.0  &  2 & -2 & 23.51 & -1.50 & 28.22 & -3.81 \\
        8.0 & 3 & -2& 29.86 & -0.98 & 33.55 & -2.50 \\
        12.0 &    4   & -3       & 42.97 & -1.08 & 45.19 & -1.50 \\
        \hline\hline
    \end{tabular}
    \caption{Frequencies $f^\mathrm{a}_n$ (real and imaginary parts in kHz) of the first three axion radial modes ($n$ denoting the mode overtone), for $\epsilon=0.1$ and different axion masses $m_a$ (in kHz),
    obtained by solving the full problem, i.e., the perturbed equations for all metric, fluid, and axion variables.}
    \label{tab:modescanp0100eps0.1radialmodesaxion}
\end{table}
\section{Summary}\label{sec:summary}
Our results show that the presence of an axion condensate inside NSs leads to significant modifications of both their equilibrium structure and oscillation spectra. 
Increasing the axion mass makes the star more compact and results in a slight shift in the frequencies of the fluid-dominated radial modes. The coupling between the fluid and the axion field introduces an efficient damping channel via axion emission, with damping timescales of order seconds for kHz-mass axions.

The presence or absence of damping in the oscillation spectrum could provide a potential observational handle on the axion mass. In particular, 
fluid modes with frequencies below $m_a$
are essentially decoupled from the axion field and exhibit almost no damping. Conversely, modes with frequencies higher than $m_a$ are strongly damped, especially in the vicinity of resonances (as illustrated in Fig.~\ref{fig:RefTableIIgamma2}).
In a hypothetical observation of radial modes, the absence of any detectable damping across the full observed spectrum would set a lower bound on the axion mass, given by the maximum measured mode frequency. On the other hand, the detection of rapidly damped modes at several frequencies would suggest that the lowest of these frequencies provides an upper bound on the axion mass. Combining the two facts, detecting undamped oscillations up to a certain frequency, beyond which no persistent modes are observed (or only heavily damped ones appear), would constrain the axion mass to be larger than the highest detected frequency. 

Although radial oscillations in NSs are unlikely to be directly observed with current instruments, the same qualitative features induced by axion couplings should also apply to nonradial oscillations, for which the prospects are more promising. 
Future work will extend this analysis to nonradial oscillations, where mode couplings and resonance phenomena are expected to play a key role. 
Nonradial modes generate GWs, the detection of which is one of the scientific goals of the upcoming generation of detectors, like the Einstein Telescope \cite{Abac_2025} and Cosmic Explorer \cite{Reitze_2019}. This would provide a unique opportunity for probing axion physics through GW observations.

\appendix

\section{Numerical procedure to compute highly damped modes}\label{app:numerical_procedure}

To numerically integrate the system \eqref{eq:Einseq_sphsymmetry_pert} we use the \texttt{solve\_bvp} function from Python's \texttt{scipy.integrate} library \cite{Virtanen_2019}, which uses a collocation algorithm to convert the differential equations into a system of algebraic equations. By iteratively refining the initial guess, it eventually converges to a solution that satisfies both the differential equations and the specified boundary conditions \cite{Kierzenka_2001}. 

However, there is a subtle point that needs to be addressed. Picking up the outgoing solution for $\delta a$ in Eq.~\eqref{eq:delta_ainf} becomes numerically challenging if the imaginary part of $k$ is large. For complex $k$, the two terms in Eq.~\eqref{eq:delta_ainf} exhibit opposing exponential behaviors, causing the ingoing term (which should vanish) to become so much smaller than the outgoing term that it makes it impossible to track, due to roundoff errors. The same numerical challenge also arises when integrating the Zerilli equation to compute the spacetime $w$-modes \cite{Kokkotas_1992, Nollert_1992, Andersson_1995}. One way to overcome this problem is by using the phase-amplitude method \cite{Andersson_1992, Andersson_1995}. In this approach, the radial coordinate $r$ can assume complex values, thus allowing us to choose a complex integration path that suppresses the exponential divergences.

Following \cite{Andersson_1995}, we apply the following variable transformations to Eq.~\eqref{eq:KGQCDaxmetric_sphsymmetry_pert_outside2}:
\begin{subequations}
    \begin{align}
        \delta a& = \left(1-\frac{2GM}{r}\right)^{-1/2}\frac{\Psi}{r},\\
        \Psi&=\frac{1}{\sqrt{Q}}\exp\left[i\int_{R}^{r} Qd r\right].
    \end{align}
\end{subequations}
 Then, we obtain the following differential equation for the phase function $Q$:
\begin{equation}\label{eq:Qeq}
        \frac{1}{2Q}\frac{d^2Q}{dr^2}-\frac{3}{4Q^2}\left(\frac{dQ}{dr}\right)^2+Q^2-W=0,
\end{equation}
where the potential $W(r)$ is given by
\begin{equation}
    W(r) \equiv\left(1-\frac{2GM}{r}\right)^{-2}\left[\left(\frac{GM}{r^2}\right)^2+k^2+\frac{2GMm_a^2}{r}\right].
\end{equation}

Extending the differential Eq.~\eqref{eq:Qeq} into the complex $r$ plane is straightforward since the potential $W$ is a known function of $r$. 
However, this equation may be inaccurate near the NS surface, where the axion background terms neglected in the derivation of \eqref{eq:KGQCDaxmetric_sphsymmetry_pert_outside2}, may become significant. 

To assess the possible errors introduced by the omission
of the axion background terms, we compare the direct integration down to the star surface with another two-step integration procedure. First, we integrate \eqref{eq:Qeq} in the complex plane from a large radial distance down to a matching point where the axion amplitude has decayed sufficiently and thus the axion background terms may still be neglected. This point is typically located at a distance of order $1/m_a$ above the NS surface. For this integration we use the simplest path: a straight line with slope $-\mathrm{Im}(k)/\mathrm{Re}(k)$ intersecting the real axis at the matching point. Then, we proceed with integrating the full system \eqref{eq:Einseq_sphsymmetry_pert_outside} along the real axis down to the NS surface. This final integration on the real line is not affected by exponential divergences, provided that $\mathrm{Im}(k/m_a)\ll1$.

We have verified that changing the location of the matching point described above and adjusting the tolerance of the boundary-value problem solver have no impact on the fluid-dominated modes, in which we typically find differences at the third significant digit in the real part of the frequencies, except for the $f_2$ mode in the case of $m_a=12$ kHz, where the uncertainty in the imaginary part is somewhat larger (about 25\%).

For the highly damped axion modes $\left(f_1^a, f_2^a\right)$, the results show some sensitivity to shifting the matching point from the star surface to $0.5/m_a$ or $1/m_a$,
with the imaginary part being more sensitive than the real part of the frequency. 
In any case, for most modes the imaginary components are accurately resolved (with errors below 1\%). The fundamental axion modes $\left(f_0^a\right)$ are 
the most sensitive to numerical inaccuracies due to the choice of matching point.
For $m_a=4$ kHz and 5 kHz, the real parts vary by about 10\%, while the damping changes by 30 \% and 10\%, respectively. At $m_a=6.0$ kHz, the estimated errors in the real part and imaginary part are about 30 \% and 40 \%, respectively. For higher axion masses, the $f_0^a$ frequencies are only reliable within a factor of 2.

To summarize, our procedure for obtaining the mode frequencies is the following:
\begin{enumerate}
    \item[1] For a chosen value of $\omega^2$, we integrate Eq.~\eqref{eq:Qeq} using the phase-amplitude method, as described above. This yields the values of the axion perturbation and its derivative at the NS surface.
    \item[2] These surface values serve as the boundary conditions for the interior problem. We solve the system of equations \eqref{eq:Einseq_sphsymmetry_pert} inside the NS, using the \texttt{solve\_bvp} function.
\end{enumerate}
The final output of this procedure is $\Delta p(R)$. The eigenmodes of the system correspond to those values of $\omega^2$ that minimize $\Delta p(R)$.

\begin{acknowledgments}
A. G. B. acknowledges the hospitality of the Institute for Nuclear Theory at the University of Washington, where part of this work was completed, and thanks S. Reddy, N. Andersson, M. Kumamoto, and T. Zhao for useful discussions. A. G. B. is supported by an ACIF 2023 fellowship, cofunded by the Generalitat Valenciana and the European Union through the European Social Fund. We acknowledge funding from the Conselleria d'Educació, Cultura, Universitats i Ocupació de la Generalitat Valenciana through the Prometeo excellence programme grant CIPROM/2022/13, as well as from the Ministerio de Ciencia e Innovaci\'on and the European Union through the grant PID2021-127495NB-I00 (MCIN/AEI/10.13039/501100011033 and EU).
\end{acknowledgments}

\bibliography{bibliography}

\end{document}